\documentclass[11pt]{article}

\usepackage{graphicx,amsfonts,latexsym,color}

\usepackage{eurosym}

\usepackage[english]{babel}
\usepackage{amssymb}
\usepackage{amsmath}
\usepackage{latexsym}
\usepackage{graphicx}

\usepackage{mathrsfs}

\newcommand{\one}{{\mathbf{1}}}
\newcommand{\tran}{^{\top}}

\newcommand{\qed}{{\hfill $\square$}}

\newcommand{\lam}{\lambda}

\newcommand{\sign}{{\mbox{\rm sgn}\,}}

\newcommand{\beq}{\begin{equation}}
\newcommand{\eeq}{\end{equation}}
\newcommand{\bea}{\begin{eqnarray}}
\newcommand{\eea}{\end{eqnarray}}
\newcommand{\beas}{\begin{eqnarray*}}
\newcommand{\eeas}{\end{eqnarray*}}
\newcommand{\ba}{\begin{array}}

\newcommand{\ea}{\end{array}}
\newcommand{\bit}{\begin{itemize}}
\newcommand{\eit}{\end{itemize}}
\newcommand{\ben}{\begin{enumerate}}
\newcommand{\een}{\end{enumerate}}
\newcommand{\dss}{\displaystyle}
\newcommand{\Real}[1]{ { {\mathbb R}^{#1} } }

\newcommand{\ped}[1]{{_{\mathrm{#1}}}}

\newtheorem{theorem}{Theorem}

\newtheorem{remark}{Remark}

\newcommand{\calE}{{\mathcal E}}
\newcommand{\calF}{{\mathcal F}}

\begin{document}
\author{Giuseppe C. Calafiore\thanks{Giuseppe C. Calafiore, Dipartimento di Automatica e Informatica,
Politecnico di Torino, Italy.
Tel.: +39-011-564.7071; Fax:
+39-011-564.7099. E-mail: {\tt giuseppe.calafiore@polito.it}
},
Laurent El Ghaoui\thanks{Laurent El Ghaoui, EECS and IEOR, UC Berkeley, CA, USA. E-mail: {\tt  elghaoui@berkeley.edu}},
Carlo Novara\thanks{Carlo Novara, Dipartimento di Automatica e Informatica,
Politecnico di Torino, Italy.
Tel.: +39-011-564.7019. E-mail: {\tt carlo.novara@polito.it}}
}

\title{Sparse Identification of Posynomial Models}

\date{}
\maketitle

\begin{abstract}

Posynomials are nonnegative combinations of monomials with possibly fractional and both positive and negative exponents. 
Posynomial models are widely used in various engineering design endeavors,  such as circuits, aerospace and structural design, mainly due to the fact that design problems cast in terms of posynomial objectives and constraints can be solved efficiently by means of a convex optimization technique known as geometric programming (GP). However, while quite a vast literature exists on GP-based design, very few contributions can yet be found on the problem of identifying posynomial models from experimental data.  Posynomial identification amounts to determining not only the coefficients of the combination, but also the exponents in the monomials, which renders the identification problem numerically hard.  In this draft, we propose an approach to the identification of multivariate posynomial models, based on the expansion on a given large-scale basis of monomials. The  model is then identified  by seeking coefficients of the combination that minimize a mixed objective, composed by a term representing the fitting error and a term inducing sparsity in the representation, which results in a problem formulation of the ``square-root LASSO'' type, with nonnegativity constraints on the variables. We propose to solve the problem via a sequential coordinate-descent scheme, which is suitable for large-scale implementations.

\vspace{.1cm}

\noindent
{\em Key Words: } Posynomial models, Identification, Sparse optimization, Square-root LASSO, Coordinate-descent methods.
\end{abstract}

\section{Introduction}

A posynomial model is defined by a function $\psi$ of the form 
\begin{equation}
\psi(w)=\sum_{i=1}^{n_{c}}c_{i}w^{\alpha_{i}}\label{eq:pp0}
\end{equation}
where $w\in\mathbb{R}_{++}^{n_{w}}$ (the positive orthant), $\psi(w)\in\mathbb{R}$,
$c_{i}\geq0$ are coefficients, $\alpha_{i}=[\alpha_{i1}\,\cdots\,\alpha_{in_{w}}]\tran\in\mathbb{R}^{n_{w}}$
are vectors of exponents with $\alpha_{ij}\in\mathbb{R}$, and $w^{\alpha_{i}}$
is defined as 
\[
w^{\alpha_{i}}\doteq\prod_{j=1}^{n_{w}}w_{j}^{\alpha_{ij}}.
\]
The term $c_{i}w^{\alpha_{i}}$ is called a {\em monomial}. Note that, while
in polynomial models the exponents $\alpha_{ij}$ are nonnegative
integers, in posynomial models these exponents may also be negative
and/or noninteger.

Posynomial models are of great importance in many fields of technology,
ranging from structural design, network flow, optimal control (see
\cite{BePh76,Wilde78}), to aerospace system design \cite{HoAb12},
circuit design \cite{bkph05,DaGiSa03,SaRaVaKa93}, antennas \cite{BaLaDoHa10}
and communication systems \cite{Chiang05}. The interest in posynomials
is motivated by the fact that they lead to computationally efficient
geometric programming models for optimal system design, see, e.g.,
\cite{gp_67,BePh76,Wilde78}.

Despite the fact that a  consistent number of papers is available in the literature
where posynomial models and geometric programming are used for design
purposes, very few works can be found to date addressing the relevant problem
of identifying a posynomial model from experimental data; see \cite{DaGiSa03}
for such an exception. Typically, the model is  assumed known (i.e.,
the coefficients $c_{i}$ and the exponents $\alpha_{ij}$ are assumed
known), and then it is processed by  geometric programming to obtain an optimal design.
However, in most real-world applications, the model is {\em not}
known a priori, and has to be identified from experimental data.

Identification of posynomial models can be performed following the
standard approach used for polynomials. In this approach, an heuristic
search finalized at finding a viable model structure, i.e., a suitable
set of exponent vectors $\left\{ \alpha_{i}\right\} $ is first carried
out. Once the exponent vector set has been chosen, the coefficients
$c_{i}$ are estimated by means of least-squares or other convex optimization
algorithms, see, e.g., \cite{SpPiLo06,PuPi07,DaGiSa03}. A critical
issue in this approach is that the model structure search may be extremely
time consuming and in most cases leads only to approximate model structures,
see \cite{MiNoAUT04}. An alternative approach is to assume (or estimate
by means of some heuristic) a value $\hat{n}_{c}$ for the basis cardinality
$n_{c}$, and then estimate $c_{i}$ and $\alpha_{i}$ by means of
nonlinear programming algorithms. However, these kind of algorithms
are non-convex and thus do not ensure convergence to the optimal parameter
estimate. A third approach, which overcomes the issues of the other
two, consists in considering an over-parametrized model and inserting
in the optimization problem a sparsity promoting term (or constraint),
given by the $\ell_{1}$-norm of the coefficient vector. This term
allows one to efficiently select the model structure and, at the same
time, to avoid the problem of overfitting. This approach is based
on the well-known LASSO (least absolute shrinkage and selection operator)
or other similar algorithms (see, e.g., \cite{Tib96,KuLoBr06,BoSePi10,NoTAC12}
for applications of the approach to identification of polynomial models).
The optimization problem is in this case convex but, due to the over-parametrization,
it typically involves a very large number of decision variables.

In this paper, we follow this latter approach: we minimize a convex
objective, defined as the sum of a regularized accuracy term based
on the $\ell_{2}$-norm of the estimation residual, and a sparsity-inducing
term given by a weighted $\ell_{1}$-norm of the coefficient vector.
We name this approach \emph{nonnegative regularized square-root LASSO}
or nnrsqrt-LASSO, since it is similar to LASSO but presents three
differences which may give advantages in terms of computational efficiency
and model regularity. The first one is to use in the objective function
an accuracy objective that is the square-root of the one used in LASSO.
With this choice, we obtain an a-priori and easily checkable sufficient condition
that, if satisfied for a certain monomial, guarantees that that monomial
will not appear in the representation (i.e., it has a null coefficient). 
This condition (called \emph{feature elimination} condition) can be
verified very efficiently, and can thus be used in a pre-optimization
phase to eliminate all the monomials which have very low relevance
in explaining the data. The second difference is to include an $\ell_{2}$
regularization in the accuracy term, allowing us to implicitly account for uncertainty
in the data, and to improve the numerical conditioning of the problem. 
The third difference
consists in using a weighted $\ell_{1}$-norm of the coefficient vector
in place of the standard $\ell_{1}$-norm. This allows for more flexibility
in problems where the entries of $c$ have different scales. Note
that in the nnrsqrt-LASSO the variables are constrained to be nonnegative,
as required for the identification of posynomial models.

In order to solve the nnrsqrt-LASSO problem, we propose a large-scale-capable
iterative algorithm based on sequential coordinate descent.

The remainder of the paper is organized as follows. In Section \ref{sec:id_pos}, the problem
of identifying a posynomial model is introduced and then formulated
in terms of a nnrsqrt-LASSO optimization problem. In Section \ref{sec:dual},
the dual formulation of this optimization problem is developed and
the feature elimination condition is derived. Section \ref{sec:univ}
shows how the univariate nnrsqrt-LASSO optimization problem can be
solved in closed form. Based on this result, in Section \ref{sec:cd},
a sequential coordinate descent scheme is proposed, allowing us
to solve the multivariate optimization problem. The computational
aspects of the proposed scheme are also discussed in this section.
Finally, in Section \ref{sec:numes}, two numerical examples are presented.
The first one regards identification of a posynomial with negative
and non integer exponents; the second one is about identification
of a posynomial model for a NACA 4412 airfoil.

\section{Identification of posynomial models}
\label{sec:id_pos}

\subsection{Model setup}
\label{pp_id}
Consider a posynomial
\begin{equation}
\psi^{o}(w)=\sum_{i=1}^{n_{c}}c_{i}^{o}w^{\alpha_{i}^{o}}\label{eq:fo}
\end{equation}
where the coefficients $c_{i}^{o}$, the exponent vectors $\alpha_{i}^{o}$
and the expansion cardinality $n_{c}$ are not known. Suppose that
a set of noise-corrupted measurements is available:
\[
\mathfrak{D}=\left\{ y(k),w(k)\right\} _{k=1}^{m}
\]
where 
\[
y(k)=\psi^{o}(w(k))+e(k)
\]
and $e(k)\in\mathbb{R}$ is a noise term. The problem considered in this
paper is to estimate from these data the unknown parameters $c_{i}^{o}$,
$\alpha_{i}^{o}$, $i=1,\ldots,n_{c}$, and the cardinality $n_{c}$.

To this end, we define an over-parametrized posynomial family
\begin{equation}
\psi(w)=\sum_{i=1}^{n}x_{i}w^{\alpha_{i}}\label{eq:model}
\end{equation}
where $n\gg n_{c}$. In real-world situations, this over-parametrization
can be obtained from the available prior information on the exponents
$\alpha_{ij}^{o}$. For example, a certain exponent may be unknown
but it can be known to be  integer and to belong to a given interval; another
one may be known to be fractional in another interval; another one can be known to be negative,
etc.

More formally, suppose that the following prior information is available
on the exponents:
\begin{equation}
\alpha_{ij}\in Q_{j}\label{eq:prior1}
\end{equation}
where $Q_{j}$ is a set of exponents which, on the basis of the available
prior information, can be considered reasonable for the variable $w_{j}$.
Then, the set of exponent vectors defining the over-parametrization
(\ref{eq:model}) can be constructed as
\[
S_{\alpha}\doteq\left\{ \alpha_{i}\right\} _{i=1}^{n}=\prod_{j=1}^{n_{w}}Q_{j}
\]
where $\prod$ denotes the Cartesian product. Note that this approach
can be adopted also if an exponent is known to belong to a continuous
(finite) interval, in which case the set $Q_{j}$ can be obtained by properly discretizing
the interval.

If the information (\ref{eq:prior1}) is correct, then $S_{\alpha}$
is guaranteed to contain the true exponent vectors:
\[
S_{\alpha}\supset S_{\alpha^{o}}\doteq\left\{ \alpha_{i}^{o}\right\} _{i=1}^{n_{c}}.
\]

\subsection{Square-root LASSO formulation of the identification problem}
Model identification is here performed by minimizing with respect to the coefficients $x_{i}$
in the expansion (\ref{eq:model})
an objective function defined as the sum of an accuracy objective
and a sparsity-promoting
term, allowing us to select, in the over-parametrized family, a parsimonious model structure. 
Define
 $y=[y(1)\,\cdots\, y(m)]^{\top}$, $x=[x_{1}\,\cdots\, x_{n}]^{\top}$,
and 
\[
\Phi=\left[\begin{array}{ccc}
w(1)^{\alpha_{1}} & \cdots & w(1)^{\alpha_{n_{w}}}\\
\vdots & \ddots & \vdots\\
w(m)^{\alpha_{1}} & \cdots & w(m)^{\alpha_{n_{w}}}
\end{array}\right].
\]
The objective we consider is of the form
\beq
f(x)\doteq\left\Vert \left[\begin{array}{c}
\Phi x-y\\
\sigma x
\end{array}\right]\right\Vert _{2}+\lambda^{\top}\left|x\right|,
\label{eq:f:obj}
\eeq
where $\sigma\geq 0$, $\lambda\in\mathbb{R}^{n}$ 
with $\lambda\geq 0$ (component-wise), and $|x|$ denotes a vector whose entries are
the absolute values of the entries in $x$.
We define, for notational
compactness, 
\[
\tilde{\Phi}\doteq\left[\begin{array}{c}
\Phi\\
\sigma I
\end{array}\right],\quad\tilde{y}\doteq\left[\begin{array}{c}
y\\
0
\end{array}\right],\quad\tilde{\phi}_{i}\doteq\left[\begin{array}{c}
\phi_{i}\\
\sigma e_{i}
\end{array}\right],
\]
where $\tilde{\phi}_{i}$, $i=1,\ldots,n$, denotes the $i$-th column
of $\tilde{\Phi}$, and $e_{i}$ is the $i$-th vector of the standard
basis of $\mathbb{R}^{n}$. The objective thus becomes
\beq
f(x)\doteq \|\tilde \Phi x -\tilde y\|_2 +\lambda^{\top}\left|x\right|.
\eeq

Note that $\lambda^{\top}\left|x\right|$ is  a weighted
$\ell_{1}$-norm. Vector $\lambda$ is thus a penalty factor which
quantifies the tradeoff between the  accuracy objective
$\|\tilde{\Phi}x-\tilde{y}\|_{2}$ and the term $\lambda^{\top}\left|x\right|$,
which is a proxy for sparsity in the solution, see \cite{Fuchs05,Tropp06,Donoho06_2,Candes06_2}.
Clearly, for $\lambda=\gamma\one$ (where $\one$
is a vector with all entries equal to one), and $\sigma=0$, the rsqrt-LASSO
problem coincides with the standard sqrt-LASSO. 
The use of the sparsity
promoting term $\lambda^{\top}\left|x\right|$ instead of the standard
term $\gamma\|x\|_{1}$  allows for more flexibility, in problems
where the entries of $x$ have different scales. 
The regularization
parameter $\sigma\geq 0$  is introduced to improve the numerical conditioning of the problem, guaranteeing
(if $\sigma >0$) that $\tilde{\Phi}$ has full rank, and that the $\ell_2$ term of the objective remains differentiable for all $x$.

We hence consider the following two optimization problems, which
we name regularized square-root LASSO (rsqrt-LASSO)
\begin{equation}
p^{*}\doteq\min_{x\in\mathbb{R}^{n}}f(x),
\label{eq:rsqrtLASSO:primal}
\end{equation}
and nonnegative regularized square-root LASSO (nnrsqrt-LASSO)
\begin{equation}
p_{+}^{*}\doteq\min_{x\in\mathbb{R}_{+}^{n}}f(x),\label{eq:nnrsqrtLASSO:primal}
\end{equation}
where $\mathbb{R}_{+}^{n}\doteq\{x\in\mathbb{R}^{n}:x\geq0\}$ (the
inequality is component-wise). The first model can be used for polynomial model identification, and the second one for posynomial model identification (the focus in this paper is on this latter case).

As already mentioned, the solutions of the optimization problems (\ref{eq:rsqrtLASSO:primal})
and (\ref{eq:nnrsqrtLASSO:primal}) tend to be sparse, i.e., to have only a few non-zero components. This important feature is produced
by the $\ell_{1}$ term, which is able to select among the large set
of monomials only those which are relevant to explain the data.
Indeed, the $\ell_{1}$-norm is the convex envelope of the $\ell_{0}$
quasi-norm, a quantity defined as the number of vector non-zero elements,
which is commonly used to measure vector sparsity. 
Minimizing
the $\ell_{1}$-norm allows one to approximately minimize the $\ell_{0}$ quasi-norm,
and thus to maximize the coefficient sparsity \cite{Fuchs05,Tropp06,Donoho06_2,Candes06_2}.
While the $\ell_{0}$ quasi-norm is non-convex and its minimization
is a NP-hard problem, the $\ell_{1}$-norm is convex and its minimization
can be performed quite efficiently. Conditions under which the $\ell_{1}$
minimization problem provides a maximally sparse solution, i.e., a
solution of the corresponding $\ell_{0}$ minimization problem, are
given, e.g., in \cite{NoTAC12}. Note that the sparsity property is
important also to allow an efficient implementation on real-time processors,
which may have limited memory and computational capacity \cite{NoFaMiAUT13}.

\begin{remark}\rm Notice that the cardinality $n$ of the set $S_{\alpha}$, and hence the dimension of the decision vector $x$, may be
very large, since it is given by the product of the cardinalities of $Q_{j}$, for $j=1,\ldots,n_w$.
For this reason, although the two previous problems are standard convex optimization problems, they may
not be practically solved using standard interior-point methods for convex optimization. Actually, in some cases, even just storing in memory the data matrix $\Phi$ may be unfeasible due to dimensionality issues.\qed

In the following sections, we describe a simple scheme for solving both the unconstrained and the constrained versions
of the regularized sqrt-LASSO problem, based on a two-phase procedure. 
In the first phase, we apply a feature elimination step to eliminate a-priori all variables that are guaranteed to be zero at optimum, thus possibly reducing the dimensionality of the problem. In the second phase, we apply a coordinate-descent scheme to the reduced problem, in order to find the optimal solution. This latter phase is based on the fact that we can find in ``closed form'' an optimal solution to the univariate restriction of the above problems. 
\end{remark}

\vspace{.2cm}
We shall assume throughout that $y\neq 0$, since for $y=0$ the optimal solution of both problems 
(\ref{eq:rsqrtLASSO:primal}), (\ref{eq:nnrsqrtLASSO:primal})   is trivially $x^*=0$.

\section{Dual  formulations and feature elimination}
\label{sec:dual}

We next derive dual formulations of the rsqrt-LASSO and nnrsqrt-LASSO problems,
and then show how a feature elimination condition is obtained from these dual formulations. 

\subsection{Dual of the rsqrt-LASSO problem}
We here derive a dual formulation for problem (\ref{eq:rsqrtLASSO:primal}).  To this end, we first recall the definition of dual norm: if $\|\cdot\|$ is a vector norm, then the corresponding dual norm is defined as
\[
\|x\|_\star \doteq \max_{\|v\|\leq 1} \, v\tran x.
\]
It is well known, for instance, that the dual of the $\ell_2$ norm is the $\ell_2$ norm itself, and that
the dual of the $\ell_\infty$ norm is the $\ell_1$ norm, and vice versa. Therefore,
\beas
\|\tilde \Phi x -\tilde y\|_2 &=& \max_{\|u\|_2\leq 1} \, u\tran (\tilde \Phi x - \tilde y).
\eeas
Also, one can readily verify that
\[
\lam\tran |x| = \sum_{i=1}^n \lam_i |x_i| = \max_{|v| \leq \lam } \, v\tran x.
\]
We can  thus rewrite problem (\ref{eq:rsqrtLASSO:primal}) as
\beas
p^*  =\min_{x\in\Real{n}} & \displaystyle{\max_{\|u\|_2\leq 1,    |v| \leq \lam}} \;
u\tran (\tilde \Phi x - \tilde y) +  v\tran x.
\eeas
Then, a standard saddle-point result (see, for instance, Sion's theorem, \cite{Komiya:88,Sion:58}), 
prescribes that we may exchange the order of min and max in the previous expression without changing the optimal value, whence
\beas
p^*  & = & \displaystyle{\max_{\|u\|_2\leq 1,   |v| \leq \lam}} \;  \displaystyle{\min_{x\in\Real{n}}} \;
u\tran (\tilde \Phi x - \tilde y)  +  v\tran x.
\eeas
Notice further that the infimum over $x\in\Real{n}$ of the term $(u\tran \tilde \Phi +  v\tran)x $ is $-\infty$, unless the coefficient
$u\tran \tilde \Phi  +  v\tran$ is zero, hence
\beas
p^*  = \displaystyle{\max_{u,v}} & 
-u\tran  \tilde y \\
\mbox{s.t.:} & \tilde \Phi\tran u   +  v = 0 \\
&\|u\|_2\leq 1 \\
& |v| \leq \lam.
\eeas
Eliminating the $v$ variable, we obtain the following  formulation
for the dual of  problem (\ref{eq:rsqrtLASSO:primal})  
 \bea
p^*  = \displaystyle{\max_{u}} & 
-u\tran  \tilde y   \label{eq:rsqrtLASSO:dual} \\
\mbox{s.t.:} & \|u\|_2\leq 1 \nonumber \\
& |\tilde \phi_i\tran u | \leq \lam_i , &  i=1,\ldots, n. \label{eq:rsqrtLASSO:dual_cinf} 
\eea

\subsection{Dual of the nnrsqrt-LASSO problem}
The derivation of the dual for the nnrsqrt-LASSO    problem (\ref{eq:nnrsqrtLASSO:primal})
follows  similar lines, noticing that,
for $x\geq 0$, we have $\lam\tran |x| = \lam\tran x$, hence 
\beas
p^*_+  & = & \displaystyle{\max_{\|u\|_2\leq 1}} \;  \displaystyle{\min_{x\geq 0}} \;
u\tran (\tilde \Phi x - \tilde y) +  \lam\tran x,
\eeas
and the infimum over $x\geq 0$ of the term $(u\tran \tilde \Phi +  \lam\tran)x $ is $-\infty$, unless 
$u\tran \tilde \Phi  +  \lam\tran\geq 0$, thus
\bea
p^*_+  = \displaystyle{\max_{u}} & 
-u\tran  \tilde y  \label{eq:rsqrtLASSO:dual+} \\
\mbox{s.t.:} & \|u\|_2\leq 1 \nonumber \\
& \tilde \phi_i\tran u   +  \lam_i \geq  0, & i=1,\ldots,n. 
 \label{eq:rsqrtLASSO:dual_cinf+} 
\eea

\subsection{Safe feature elimination}
\label{sec:safelim}
In this section we analyze the dual formulations of problems (\ref{eq:rsqrtLASSO:primal}), (\ref{eq:nnrsqrtLASSO:primal}), in order to derive
a simple sufficient condition that permits to predict when an entry
$x_i$ is zero  at optimum, and hence to
eliminate a priori some features (i.e., columns of $\tilde \Phi$) from the problem.
This type of condition, first introduced by \cite{ElgViRa:12} in the context of the standard LASSO problem, is named {\em safe feature elimination}.
Observe that
\beas
\max_{ \|u\|_2\leq 1}\, |\tilde \phi_i\tran u | &=& \|\tilde\phi_i\|_2 = 
\left\|\left[\ba{c}\phi_i \\ \sigma e_i \ea\right]\right\|_2.
\eeas
Therefore, if for some $i\in\{1,\ldots,n\}$ it holds that
\[
\left\|\left[\ba{c}\phi_i \\ \sigma e_i \ea\right]\right\|_2^2 = 
\|\phi_i\|_2^2 + \sigma^2 < \lam_i^2
\]
then the corresponding constraint
in (\ref{eq:rsqrtLASSO:dual_cinf}), 
as well as in (\ref{eq:rsqrtLASSO:dual_cinf+}),
will certainly be satisfied with strict inequality, that is, it will be {\em inactive}
at the optimum.
This means that  it can be safely eliminated from the dual optimization problem, without changing the optimal objective value. Defining
\[
\calF(\lam) \doteq \{i:\,  \|\phi_i\|_2^2 +\sigma^2 \geq  \lam_i^2,\;  i=1,\ldots, n\},
\]
we thus have that
 \bea
p^*  = \displaystyle{\max_{u}} & 
-u\tran  \tilde y   \label{eq:rsqrtLASSO:dual_red} \\
\mbox{s.t.:} & \|u\|_2\leq 1 \nonumber \\
& |\tilde \phi_i\tran u  | \leq \lam_i , &  i\in\calF(\lam) \label{eq:rsqrtLASSO:dual_cinf_red} \nonumber ,
\eea
which is the dual of the ``reduced'' primal problem
\bea
p^*  = \min_{\xi} &
\|\tilde \Phi_{\calF(\lam)} \xi  - \tilde y 
\|_2
 + \lam\tran  |\xi|,
\label{eq:rsqrtLASSO:primal_red}
\eea
where $\tilde \Phi_{\calF(\lam)}$ is a matrix containing by columns vectors $\tilde \phi_i$, $i\in\calF(\lam)$,
and $\xi$ is a decision variable vector, having dimension equal to the cardinality of $\calF(\lam)$.
In other words, the features $x_i$ in the primal problem (\ref{eq:rsqrtLASSO:primal})
corresponding to indexes $i$ in the set $\calE(\lam)$  complementary to $\calF(\lam)$, defined as
\[
\calE(\lam) \doteq \{i:\,  \|\phi_i\|_2^2 +\sigma^2  < \lam_i^2,\;  i=1,\ldots, n\},
\]
are certainly zero at the optimum, that is
\beq
\|\phi_i\|_2^2 + \sigma^2 < \lam_i^2  \quad \Rightarrow \quad x_i^* = 0.
\label{eq:feature-elim}
\eeq
Similarly,  we have that
 \bea
p^* _+ = \displaystyle{\max_{u}} & 
-u\tran  \tilde y   \label{eq:nnrsqrtLASSO:dual_red} \\
\mbox{s.t.:} & \|u\|_2\leq 1   \nonumber \\
& \tilde \phi_i\tran u   + \lam_i \geq 0, &  i\in\calF(\lam) \label{eq:nnrsqrtLASSO:dual_cinf_red} \nonumber ,
\eea
is the dual of  the ``reduced'' primal problem
\bea
p^*_+  = \min_{\xi\geq 0} & \|\tilde \Phi_{\calF(\lam)} \xi  - \tilde y \|_2+ \lam\tran  |\xi|.
\label{eq:nnrsqrtLASSO:primal_red}
\eea

\subsubsection{When is $x=0$ optimal?}
Point  $x=0$ is optimal 
 for problem (\ref{eq:rsqrtLASSO:primal}) if and only if $p^* = \|\tilde y\|_2$, which is equivalent to $u = -\tilde y/\|\tilde y\|_2$  being optimal (hence feasible) for the dual problem. This happens if and only if
\[
|\tilde \phi_i\tran \tilde y| \leq \lam_i {\|\tilde y\|_2},\quad  i=1,\ldots,n,
\]
that is, since $\tilde \phi_i\tran \tilde y = \phi_i\tran y$, $\|\tilde y\|_2=\|y\|_2$, if an only if
\[
| \phi_i\tran  y| \leq \lam_i {\| y\|_2},\quad  i=1,\ldots,n.
\]
Similarly, point  $x=0$ is optimal 
 for problem (\ref{eq:nnrsqrtLASSO:primal}) if and only if $p^*_+ = \|\tilde y\|_2$, which is equivalent to $u = -\tilde y/\|\tilde y\|_2$ being optimal (hence feasible) for the dual problem, which happens if and only if
\[
\tilde \phi_i\tran \tilde y \leq \lam_i {\|\tilde y\|_2}, \quad  i=1,\ldots,n,
\]
or, equivalently,
\[
\phi_i\tran y \leq \lam_i {\| y\|_2}, \quad  i=1,\ldots,n.
\]

\section{Univariate solution of rsqrt-LASSO and  nnrsqrt-LASSO}
\label{sec:univ}

Consider the following  rsqrt-LASSO problem with a single scalar variable $x$:
\[
\min_{x\in\Real{}}\,  f(x)\doteq
\left\|\left[\ba{c}
\phi x - y \\
\sigma e x - \xi
\ea\right]\right\|_2 +  \lam |x|,
\]
where $\lam,\sigma\geq 0$,  $\phi\in\Real{m}, y\in\Real{m}$, $\xi\in\Real{n}$ are given, and $e$ is a vector of all zeros, except for an entry in generic position $i$, which is equal to one, and correspondingly we postulate that
$\xi_i=0$, thus it holds that $e\tran\xi=0$.
We set for convenience 
\beq
\tilde\phi \doteq  \left[\ba{c}
\phi \\ \sigma e
\ea\right], \quad \tilde y \doteq \left[\ba{c}
y \\ \xi
\ea\right].
\label{eq:tildephiy}
\eeq
Thus, the problem rewrites to
\beq
\min_{x\in\Real{}}\,  f(x)\doteq
\|
\tilde \phi x - \tilde y \|_2 +  \lam |x|.
\label{eq:univariate_sqrtLASSO}
\eeq
We  assume that $\tilde y \neq 0$ and
$\tilde \phi\neq 0$,
otherwise the optimal solution is simply $x=0$. Let us define
\[
x\ped{ls} \doteq 
\frac{\tilde\phi\tran \tilde y}{\|\tilde \phi\|_2^2} =
\frac{\phi\tran y}{\|\phi\|_2^2+\sigma^2},
\label{eq:xls}
\]
which corresponds to the solution of the problem for $\lam = 0$.
The following theorem holds.

\begin{theorem} 
\label{prop:univariate_sqrtLASSO}
Consider problem (\ref{eq:univariate_sqrtLASSO}), with $\tilde y\neq 0$, $\tilde \phi\neq 0$, $\lam\geq 0$.
\ben
\item $x^* = 0$ is an optimal solution for (\ref{eq:univariate_sqrtLASSO}) if and only if
\[
|\tilde \phi\tran \tilde y| \leq \lam \|\tilde y\|_2
\]
(notice, in particular, that if $\|\tilde \phi\|_2 \leq \lam$, then the above condition is certainly satisfied, hence $x^*=0$).
\item If $|\tilde \phi\tran \tilde y| > \lam \|\tilde y\|_2$ (hence  $\|\tilde \phi\|_2 > \lam $), then 
the optimal solution of (\ref{eq:univariate_sqrtLASSO}) is given by
\beq
x^* = x\ped{ls} -  \sign (x\ped{ls}) \frac{\lam }{\|\tilde \phi\|_2^2}  \sqrt{\frac{\|\tilde \phi\|_2^2 \|\tilde y\|_2^2 - (\tilde \phi\tran \tilde y)^2}
{\|\tilde \phi\|_2^2- \lam^2}}.
\label{eq:univariate_sqrtLASSO_nzsol}
\eeq
\een
\end{theorem}

\noindent
{\bf Proof.}
The problem is convex but nonsmooth, hence we write the optimality conditions in terms of the subdifferential of the objective:
\[
0\in \partial f(x)  =  \partial \|\tilde \phi x - \tilde y\|_2+ \lam \partial |x|,
\]
where
\beas
\partial  \|\tilde \phi x - \tilde y\|_2 &=& \left\{\displaystyle{ \ba{ll} 
\displaystyle{ \frac{\tilde \phi\tran (\tilde \phi x -  \tilde y)}{\|\tilde \phi x - \tilde y\|_2 }} & \mbox{if } \tilde \phi x - \tilde y\neq 0 \\
\{\tilde \phi\tran g:\; \|g\|_2\leq 1\} &  \mbox{if } \tilde \phi x - \tilde y= 0 ,
\ea}\right. \\
\partial  |x| &=& \left\{\ba{ll} 
\sign(x) & \mbox{if } x \neq 0 \\
\{v:\;  |v| \leq 1\}  & \mbox{if } x= 0.
\ea\right.
\eeas
For point 1.\ we thus check under what conditions $0$ is contained in the subdifferential of $f$ at $x=0$, that is
\[
\begin{array}{c}
x^{*}=0\mbox{ is optimal}\\
\Updownarrow\\
0\in\partial f(0)=\left\{ \frac{\tilde{\phi}\tran\tilde{y}}{\|\tilde{y}\|_{2}}+\lam v,\;|v|\leq1\right\} .
\end{array}
\]
Since  the term $\lam v$ may take any value in the interval $[-\lam,\, \lam]$, it follows that the above condition is satisfied if and only if
${|\tilde \phi\tran \tilde y|}/{\|\tilde y\|_2} \leq \lam$, which proves the first part of the theorem.
Also, since by the Cauchy-Schwartz inequality it holds that
\[
|\tilde \phi\tran \tilde y| \leq \|\tilde \phi\|_2 \|\tilde y\|_2 ,
\]
it is clear that $\|\tilde \phi\|_2\leq \lam$ implies $|\tilde \phi\tran \tilde y| \leq \lam\|\tilde y\|_2 $, hence the optimal solution is certainly zero when $\|\tilde \phi\|_2\leq \lam$.

\vspace{.2cm}
Consider next the case when the optimal solution is nonzero, i.e., when $|\tilde \phi\tran \tilde y| >\lam \|\tilde y\|_2$,
thus $\|\tilde \phi\|_2 > \lam$. We initially assume for simplicity that $\tilde \phi$ and $\tilde y$ are not collinear, so that
$\tilde \phi x - \tilde y\neq 0$ for all $x$; later we show that the derived solution is still valid if this assumption is lifted.
With this assumption, and since $x\neq 0$, we have that
\[
x \mbox{ is optimal}\quad \Leftrightarrow \quad
0 =\partial f(x) = \frac{\tilde \phi\tran (\tilde \phi x -  \tilde y)}{\|\tilde \phi x - \tilde y\|_2 } + \lam\,\sign(x),
\]
that is, since $\|\tilde \phi x - \tilde y\|_2\neq 0$, for
\beq
\tilde \phi\tran (\tilde \phi x -  \tilde y) = -\lam  \|\tilde \phi x - \tilde y\|_2\sign(x).
\label{eq:univariate_sqrtLASSO_rooteq}
\eeq
All solution to this equation are also solutions of the squared equation
\beq
(\tilde \phi\tran \tilde \phi x - \tilde \phi\tran \tilde y)^2 = \lam^2  \|\tilde \phi x - \tilde y\|_2^2,
\label{eq:univariate_sqrtLASSO_rooteqsq}
\eeq
which is a quadratic equation in $x$, equivalent to:
\[
\|\tilde \phi\|_2^2(\|\tilde \phi\|_2^2-\lam^2) x^2 -2\tilde \phi\tran \tilde y (\|\tilde \phi\|_2^2-\lam^2) x+ (\tilde \phi\tran \tilde y)^2-\lam^2\|\tilde y\|_2^2 = 0.
\]
The roots of this equation are in
\[
x_{\pm} = x\ped{ls} \pm \sqrt{  x\ped{ls}^2 - \frac{(\tilde \phi\tran \tilde y)^2-\lam^2\|\tilde y\|_2^2}
{\|\tilde \phi\|_2^2(\|\tilde \phi\|_2^2-\lam^2)}.
}
\]
Observe that the term under the square root is nonnegative, since
\beas
\delta \doteq x\ped{ls}^2 - \frac{(\tilde \phi\tran \tilde y)^2-\lam^2\|\tilde y\|_2^2}
{\|\tilde \phi\|_2^2(\|\tilde \phi\|_2^2-\lam^2) }&=&
\frac{(\tilde \phi\tran \tilde y)^2}{\|\tilde \phi\|^4} - \frac{(\tilde \phi\tran \tilde y)^2-\lam^2\|\tilde y\|_2^2}
{\|\tilde \phi\|_2^2(\|\tilde \phi\|_2^2-\lam^2)}  \\
&=& \frac{\lam^2}{\|\tilde \phi\|_2^2}\cdot \frac{\|\tilde \phi\|_2^2\|\tilde y\|_2^2 - (\tilde \phi\tran \tilde y)^2}
{\|\tilde \phi\|_2^2(\|\tilde \phi\|_2^2-\lam^2) },
\eeas
where, under the conditions of point 2., $\|\tilde \phi\|_2^2-\lam^2 > 0$, and 
$\|\tilde \phi\|_2^2\|\tilde y\|_2^2 - (\tilde \phi\tran \tilde y)^2 \geq 0$, by the Cauchy-Schwartz inequality.
Further,  $\delta\geq 0$ is smaller in magnitude than $x\ped{ls}^2$, since
the condition $|\tilde \phi\tran \tilde y| >\lam \|\tilde y\|_2$
implies that $x\ped{ls}^2 - \delta > 0 $.
It follows that the sign of $x_\pm = x\ped{ls} \pm \sqrt{\delta}$
is the same sign of $x\ped{ls}$ (since adding $\pm \sqrt{\delta}$ to $x\ped{ls}$
cannot change its sign).
Then, plugging $x\gets x_\pm$ into equation (\ref{eq:univariate_sqrtLASSO_rooteq}), we have 
the left-hand side
\[
\|\tilde \phi\|_2^2 x_\pm  - \tilde \phi\tran \tilde y  = 
\|\tilde \phi\|_2^2 (x\ped{ls} \pm \sqrt{\delta}) - \tilde \phi\tran \tilde y =
\pm \sqrt{\delta}
\]
and the right-hand side
\[
-\lam  \|\tilde \phi x_\pm - \tilde y\|_2\sign(x_\pm) =
-\lam  \|\tilde \phi x_\pm - \tilde y\|_2\sign(x\ped{ls}).
\]
Thus, sign consistency is obtained by choosing the solution with ``+'' when $x\ped{ls}$ is negative, and
with ``-'' when $x\ped{ls}$ is positive. In conclusion, the unique solution to eq.\ (\ref{eq:univariate_sqrtLASSO_rooteq}) is given by
\[
x^* = x\ped{ls} -\sign (x\ped{ls}) \frac{\lam}{\|\tilde \phi\|_2^2}\sqrt{\frac{\|\tilde \phi\|_2^2\|\tilde y\|_2^2 - (\tilde \phi\tran \tilde y)^2}
{\|\tilde \phi\|_2^2-\lam^2 }},
\]
which is the expression we wished to prove.

It only remains to be proved that the above expression is still valid also when $\tilde y$ and $\tilde \phi$ are collinear.
In this case, since $\|\tilde \phi\|_2^2\|\tilde y\|_2^2  = (\tilde \phi\tran \tilde y)^2$, eq.\ (\ref{eq:univariate_sqrtLASSO_nzsol}) gives $x^* = x\ped{ls}$, and
we have that $\tilde \phi  x^* - \tilde y = 0$. Let us check that this solution is indeed optimal. The subdifferential of $f$
at $x^*\neq 0$ such that $\tilde \phi  x^* - \tilde y = 0$ is
\[
\partial f(x^*) = \{\tilde \phi\tran g + \lam \,\sign(x^*), \; \|g\|_2\leq 1\},
\]
and we see that $0\in \partial f(x^*) $ if $\|\tilde \phi\|_2 \geq \lam$, which is indeed the condition under which
the expression (\ref{eq:univariate_sqrtLASSO_nzsol}) for $x^*$ holds.
\qed

\subsection{Univariate solution of nnrsqrt-LASSO}
The solution of the univariate nnrsqrt-LASSO problem in the scalar variable $x$
\beq
\min_{x\geq 0}\,  f(x)\doteq
\|\tilde \phi x - \tilde y \|_2 +  \lam |x|,
\label{eq:univariate_sqrtLASSO+}
\eeq
can be readily obtained from the solution of the corresponding unconstrained problem (\ref{eq:univariate_sqrtLASSO}), by the following reasoning. Since (\ref{eq:univariate_sqrtLASSO+}) is a convex optimization problem in one variable and one linear inequality constraint, its optimal solution is either on the boundary of the feasible set (in this case, at $x=0$), or it coincides with the solution of the unconstrained version of the problem.
Thus, we solve the unconstrained problem  (\ref{eq:univariate_sqrtLASSO}): if this solution is nonnegative, then it is also the optimal solution to  (\ref{eq:univariate_sqrtLASSO+}); if it is negative, then the optimal solution to (\ref{eq:univariate_sqrtLASSO+}) is $x=0$. Since the sign of the solution of  (\ref{eq:univariate_sqrtLASSO}) is simply the sign of $\tilde \phi\tran \tilde y$, we can state the following theorem.

\begin{theorem} 
\label{prop:univariate_sqrtLASSO+}
Consider problem (\ref{eq:univariate_sqrtLASSO+}), with $\tilde y\neq 0$, $\tilde \phi\neq 0$, $\lam\geq 0$.
\ben
\item $x^* = 0$ is an optimal solution for (\ref{eq:univariate_sqrtLASSO+}) if and only if
\[
\tilde \phi\tran \tilde y \leq \lam \|\tilde y\|_2.
\]
\item Otherwise, the optimal solution of (\ref{eq:univariate_sqrtLASSO+}) is given by
\beq
x^* = x\ped{ls} - \frac{\lam}{\|\tilde \phi\|_2^2}\sqrt{\frac{
\| \tilde \phi\|_2^2 \| \tilde y \|_2^2 - (\tilde  \phi\tran \tilde y )^2}
{\|\tilde \phi\|_2^2  -\lam^2 }}
.
\label{eq:univariate_sqrtLASSO+_nzsol}
\eeq
\een
\end{theorem}

\begin{remark}\rm
For the specific structure of $\tilde\phi$ and $\tilde y$ in (\ref{eq:tildephiy}), we have that
\[
\|\tilde \phi\|_2^2=\|\phi\|_2^2+\sigma^2, \quad \tilde \phi\tran \tilde y = \phi\tran y, \quad
\|\tilde y \|_2^2 = \|y \|_2^2 + \|\xi\|_2^2,
\]
and the solutions in theorems~\ref{prop:univariate_sqrtLASSO} and \ref{prop:univariate_sqrtLASSO+} can be expressed accordingly in terms of
$\phi\tran y$, 
$\|\phi\|_2$, $\|y \|_2$,  $\|\xi\|_2$, and $\sigma$, $\lam$.
In particular, the condition for $x=0$ being optimal becomes
\[
|\phi\tran y| \leq \lam \sqrt{\|y\|_2^2 + \|\xi\|_2^2},
\]
which, in particular, is satisfied if $\|\phi\|_2^2+\sigma^2\leq \lam^2$.\newline 
 Notice further that $\tilde\phi x - \tilde y\neq 0$  for $x=0$, since we assumed $\tilde y \neq 0$, and that,
 for $\sigma>0$, $\tilde\phi x - \tilde y\neq 0$ also for $x\neq 0$, since the $i$-th entry of $\xi$ is zero by definition.
Therefore, for $\sigma>0$, the $\ell_2$-norm part of the objective is always nonzero, and hence differentiable.\qed
\end{remark}

\section{Sequential coordinate descent scheme}
\label{sec:cd}

We next outline a sequential coordinate-descent scheme for the rsqrt-LASSO problem (\ref{eq:rsqrtLASSO:primal}).
Suppose all variables $x_j$, $j\in\{1,\ldots,n\}\setminus i$, are fixed to some numerical values, and we wish
to minimize the objective in (\ref{eq:rsqrtLASSO:primal}) with respect to the scalar variable $x_i$. We have that
\beas
f_i(x_i) &\doteq & \|\sum_{j=1}^n 
\tilde  \phi_j  x_j - 
\tilde y\|_2 + \sum_{j=1}^n \lam_j|x_j| \\
&=& 
 \|\tilde \phi_i x_i - 
\tilde y(i) \|_2 + \lam_i|x_i| + \sum_{j\neq i} \lam_j|x_j|,
\eeas
where we defined 
$
\tilde y(i) \doteq \tilde y - \sum_{j\neq i} \tilde \phi_j x_j
$.
We thus have that
\[
\begin{array}{l}
x_i^* \doteq \arg \min_{x_i} \; f_i(x_i) \\ = \arg \min_{x_i} \;  \|\tilde \phi_i x_i - 
\tilde y(i) \|_2 + \lam_i|x_i| ,
\end{array}
\]
where  the minimizer $x_i^*$ is readily computed by applying Theorem~\ref{prop:univariate_sqrtLASSO}.

A
 sequential coordinate-descent scheme  works by updating the variables $x_i$ sequentially, according to
the above univariate minimization criterion. The scheme of the algorithm is as follows.
\ben
\item Initialize $x^{(0)} = 0$ (an $n$-vector of zeros), $k=1$;
\item For $i=1,\ldots,n$, let
\[
\begin{array}{r}
x^{(k)}_i = \arg\min_{x_i} f(x_1^{(k)},\ldots,x_{i-1}^{(k)},x_i, \\ x_{i+1}^{(k-1)},\ldots, x_{n}^{(k-1)}) ;
\end{array}
\]
\item If stopping criterion is met, finish and return $x^{(k)}$, else set $k\gets k+1$, and goto 2.
\een
The detailed data management involved in applying this scheme to our specific problem is described in Section~\ref{sec:cd-iterate}.

\begin{remark}\rm 
As a stopping criterion, one may use a standard check on sufficient progress in objective reduction, or the approach described in Section~\ref{sec:dualbound}, based on the evaluation of a lower bound on the duality gap. \qed
\end{remark}

\begin{remark}\rm 
Observe that, due to Theorem~\ref{prop:univariate_sqrtLASSO}, all variables $x_i$ for which
$\|\tilde \phi_i\|_2  \leq \lam_i$ are {\em never} updated by the algorithm, i.e., they remain fixed at their initial zero value.
The inner loop on $i$ can thus be sped up by considering only the indices $i$ such that
$\|\tilde \phi_i\|_2 > \lam_i$, which can be determined a priori (feature elimination). \qed
\end{remark}

\begin{remark}\rm 
The same coordinate-descent scheme can be
used also for solving the nnrsqrt-LASSO problem  (\ref{eq:nnrsqrtLASSO:primal}), by using the result in Theorem~\ref{prop:univariate_sqrtLASSO+}
for updating the $i$-th coordinate. \qed
\end{remark}

Convergence of the proposed scheme is established in the following theorem, which is a
direct consequence of a result in \cite{Tseng:01}.

\begin{theorem}[Convergence]
For $\sigma>0$, $y\neq 0$,  the sequential coordinate descent algorithm converges
to an optimal point,
for both the rsqrt-LASSO and the nnrsqrt-LASSO problems.
\end{theorem}

\noindent
{\bf Proof.}
	The function $f(x)$ in (\ref{eq:f:obj}) that we minimize using coordinate descent is convex and composite:
	\[
	f(x) = f_0(x) + \sum_{i=1}^n \psi_i(x_i),
	\]
	where $\psi_i$ are convex and nonsmooth. 
	In the unconstrained case, we have $\psi_i(x_i)=\lam_i |x_i|$. The constrained case, where $x_i\geq 0$, can  also be tackled as an unconstrained one, by considering
	$\psi_i(x_i) = \lam_i |x_i| + I_+(x_i)$, where $I_+(x_i)$ is equal to zero if
	$x_i\geq 0$ and it is $+\infty$ otherwise. 	
	Further, the function $f_0(x)=\|\tilde\Phi x-\tilde y\|_2$ is convex and, for $\sigma>0$ and $y\neq 0$, 
	it is differentiable over all $x\in \Real{n}$.
	Since the objective we minimize satisfies the hypotheses of
	Theorem~5.1 in \cite{Tseng:01}, convergence 
	of the sequential coordinate descent algorithm to an optimal point
	 is guaranteed for both the rsqrt-LASSO and the nnrsqrt-LASSO problems.
	  \qed

\subsection{Dual-bound based stopping criterion}
\label{sec:dualbound}
Inspecting the primal and dual problems (\ref{eq:rsqrtLASSO:primal}), (\ref{eq:rsqrtLASSO:dual}), we see that if
$x^*$ is primal optimal, then the dual-optimal variable $u$ must be
\[
u^* = \frac{\tilde \Phi x^* - \tilde y}{\|\tilde \Phi x^* - \tilde y\|_2}.
\]
This suggests considering,
for the candidate solution $x^{(k)}$ at iteration $k$ of the algorithm,  an associated vector
\[
u^{(k)}  \doteq \alpha^{(k)} 
\tilde u^{(k)},\quad    
 \tilde u^{(k)}\doteq \frac{\tilde \Phi x^{(k)} - \tilde y }{\|\tilde \Phi x^{(k)} - \tilde y \|_2},
\]
where
\[
\alpha^{(k)}
 = \left\{ \ba{ll} 
1 & \mbox{if } |\tilde \Phi\tran  \tilde u^{(k)}| \leq \lam \\
{\dss \min_{i} \frac{\lam_i}{|\tilde \phi_i\tran  \tilde u^{(k)}| } }& \mbox{otherwise. }
\ea\right.
\]
Such $u^{(k)}$ is, by construction, feasible for the dual problem  (\ref{eq:rsqrtLASSO:dual}), hence 
\[
d^{(k)} \doteq -\tilde y\tran u^{(k)} = \alpha^{(k)}  \frac{\|\tilde y\|_2^2-\tilde y\tran\tilde \Phi x^{(k)}}{\|\tilde \Phi x^{(k)} - \tilde y \|_2}
\]
is a lower bound on the primal optimal value $p^*$, that is $d^{(k)} \leq p^* \leq p^{(k)}$,
where $p^{(k)} \doteq f(x^{(k)})$.
As $x^{(k)}$ converges to $x^*$, $u^{(k)} $ should converge to $u^*$ and $d^{(k)}$ to $p^*$.
Hence, if at iteration $k$ it holds that
\[
p^{(k)} - d^{(k)} \leq \epsilon,
\]
we can terminate the algorithm with a solution $x^{(k)}$ that guarantees $\epsilon$-suboptimality.

An analogous approach can be followed for determining a dual lower bound for the 
nnrsqrt-LASSO problem (\ref{eq:nnrsqrtLASSO:primal}). 
The only difference is in the choice of $\alpha^{(k)}$, which is now given by
\[
\alpha^{(k)}
 = \left\{ \ba{ll} 
1 & \mbox{if } \tilde \Phi\tran  \tilde u^{(k)} \geq - \lam \\
{\dss \min_{\{i:\, \tilde \phi_i\tran  \tilde u^{(k)} < -\lam_i\}} \frac{\lam_i}{|\tilde \phi_i\tran  \tilde u^{(k)}|} }& \mbox{otherwise. }
\ea\right.
\]

\subsection{Data management and cost per iteration}
\label{sec:cd-iterate}
We next analyze in more detail 
the data management and
the computational cost per iteration of the coordinate-descent scheme.

\subsubsection{Variable update}
Suppose we have a current value of $x$ 
and we want to update the $i$-th coordinate of $x$. Suppose further that the following quantities are available:
\beas
h &\doteq & \tilde\Phi \tran r  \\
c &\doteq & \|r\|_2^2,
\eeas
where 
\[
r \doteq \tilde\Phi x -\tilde y
\]
is the current value of the residual vector (as we shall see, we do not need to store $r$: only $h$ and $c$ need be updated).
We set up the univariate minimization problem
\[
\min_{z}\; \|\tilde\phi_i z - \tilde y(i)\|_2 + \lam_i |z|,
\] 
where 
\beas
\tilde y(i) &=& \tilde y -\sum_{j\neq i} \tilde\phi_j x_j = \tilde \phi_i x_i - (\tilde\Phi x-\tilde y) \\
&=& \tilde \phi_i x_i - r.
\eeas
Notice that  all we need in order to compute the optimal coordinate $z^*$, by applying Theorem~\ref{prop:univariate_sqrtLASSO} (or Theorem~\ref{prop:univariate_sqrtLASSO+}, in the nonnegative constrained case) is the following data:
\beas
\tilde\phi_i\tran \tilde y(i) &=& \|\tilde \phi_i\|_2^2 x_i - h_i\\
\|\tilde y(i) \|_2^2 &=& \|\tilde \phi_i\|_2^2 x_i^2 + c - 2x_i h_i.
\eeas
 Therefore, we find the optimal $z^*$, and we update the solution $x$ to
\[
x_+ = x + e_i (z^*-x_i) =  x + e_i \delta_i,
\]
where $\delta_i \doteq z^*-x_i$. Also, we update the data necessary for the next iteration. Since
\beas
r_+ &\doteq &  \tilde \Phi x_+ -\tilde y  = r + \tilde\phi_i\delta_i,
\eeas
we have that
\beas
c_+ &\doteq & \|r_+\|_2^2 = c + \|\tilde\phi_i\|_2^2\delta_i^2 + 2\delta_i h_i, \\
h_+ &\doteq & \tilde\Phi \tran r_+ =  h + \tilde\Phi \tran \tilde\phi_i\delta_i.
\eeas
Then, we let $i\gets i+1$,  $h\gets h_+$, $c\gets c_+$, $x\gets x_+$ and iterate.
The whole process is initialized with $x = 0$, $h=-\tilde\Phi\tran\tilde y$, $c=\|\tilde y\|_2^2$.

\subsubsection{Storage and computational cost per iteration}
Let us define the {\em kernel} matrix $\tilde K\in\Real{n,n}$ and the projected response vector $ q\in\Real{n}$
\beas
\tilde K &\doteq & \tilde\Phi\tran \tilde \Phi  = K + \sigma^2 I_n \\
 q &\doteq & \tilde\Phi\tran \tilde y  = \Phi\tran y,
\eeas
where
$
K \doteq \Phi\tran \Phi
$.
Initialization of the  coordinate descent method  requires
$h = -q$, and $c=\|y\|_2^2$, as described previously.

For updating the $i$-th variable, the method does not necessarily need to store or access the whole kernel matrix
$\tilde K$. Indeed, computing the $i$-th optimal update just requires access to $\|\tilde \phi_i\|_2^2 = \tilde K_{ii}$, and $O(1)$ operations. Then, the update of the $h$ vector requires access to the $i$-th column
of $\tilde K$, and then $n$ operations for computing $h_+$.

The storage requirement of the method is thus essentially given by keeping in memory  $h\in\Real{n}$
and $x\in\Real{n}$, so it is $O(n)$, if $\tilde K$ is not stored. Evaluating the $i$-th column of the kernel matrix requires $O(mn)$ operations, unless the values of the kernel can be obtained directly (i.e., without actually performing the inner products $\phi_i\tran \phi_j$), as it is the case, for instance, for polynomial kernels.

\section{Numerical examples}
\label{sec:numes}

\subsection{Example 1: posynomial with negative and non-integer exponents}

\label{sub:toy_example}

As a first numerical experiment, we have considered the problem of
identifying the posynomial function $\psi^{o}:\mathbb{R}^{3}\rightarrow\mathbb{R}$
defined as
\begin{equation}
\psi^{o}(w)=w_{2}^{1.5}w_{3}^{3}+2w_{1}^{2}w_{3}^{-1}+3w_{2}^{3.2}+4w_{1}^{0.5}w_{2}^{-2}w_{3},\label{eq:psi_es2}
\end{equation}
which contains monomials with negative
and non-integer exponents. 

A set $\mathfrak{D}=\left\{ y(k)=\psi^{o}(w(k))+e(k),w(k)\right\} _{k=1}^{m}$
of $m=600$ input-output data points has been generated from \eqref{eq:psi_es2},
for randomly chosen values of $w_{i}$ in the interval $[0.2,3.2]$,
for $i=1,2,3$. The sequence $e(k)$ has been generated as a Gaussian
noise with zero mean and a noise-to-signal standard deviation ratio
of 1\%. 

The exponent sets 
\[
\begin{array}{l}
Q_{1}=\{0,0.5,\ldots,3.5,4\}\\
Q_{2}=\{-2,-1.9,\ldots,3.9,4\}\\
Q_{3}=\{-1,0,\ldots,3,4\}
\end{array}
\]
have been assumed. For $m=600$ and for this exponent set, $\Phi$
results to be a $600\times3294$ matrix.

\begin{figure}
\centering\includegraphics[scale=0.5]{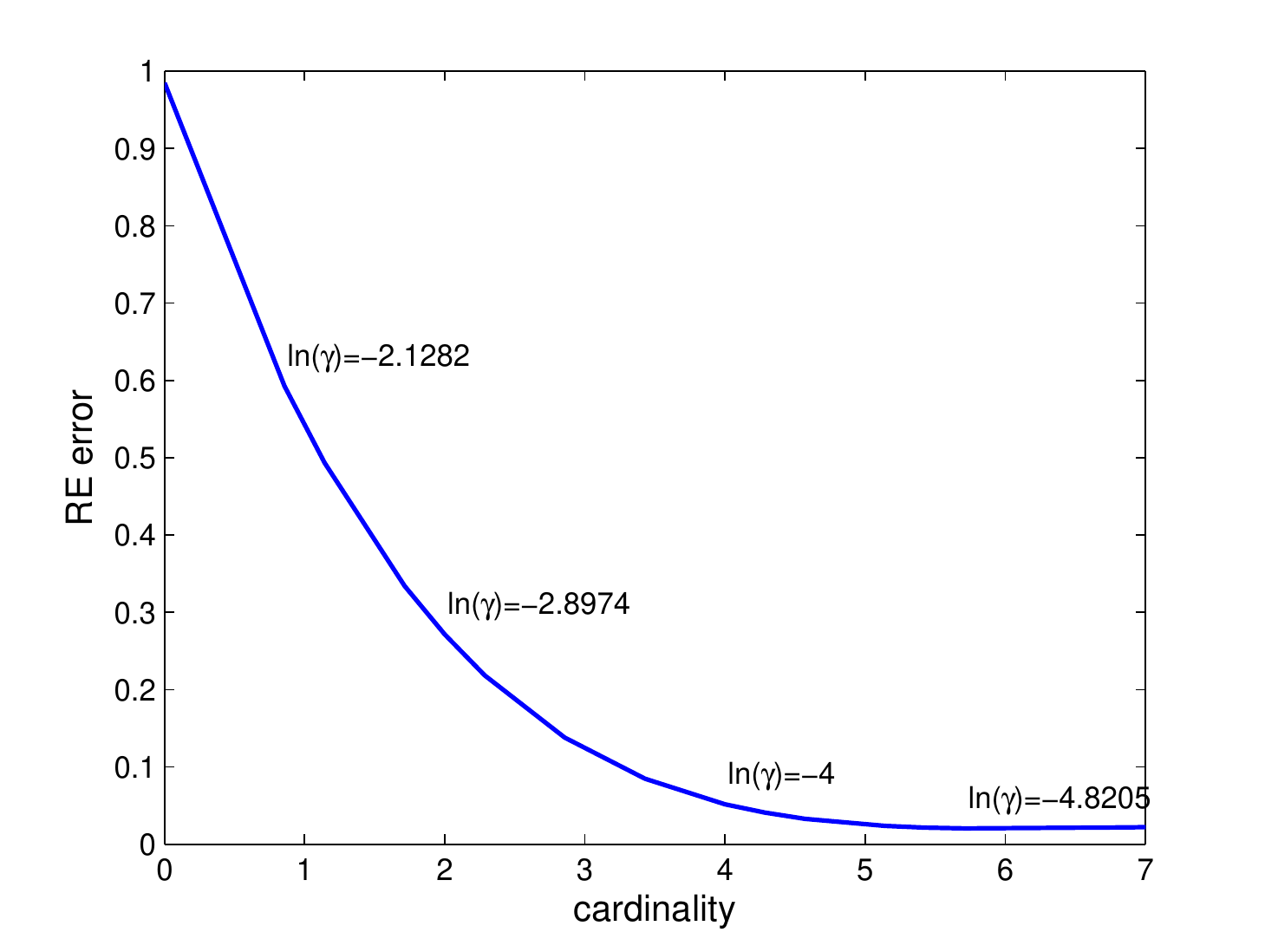}

\caption{Example 1. Pareto trade-off curve.\label{pareto:figure}}
\end{figure}

\begin{figure}
\centering\includegraphics[scale=0.5]{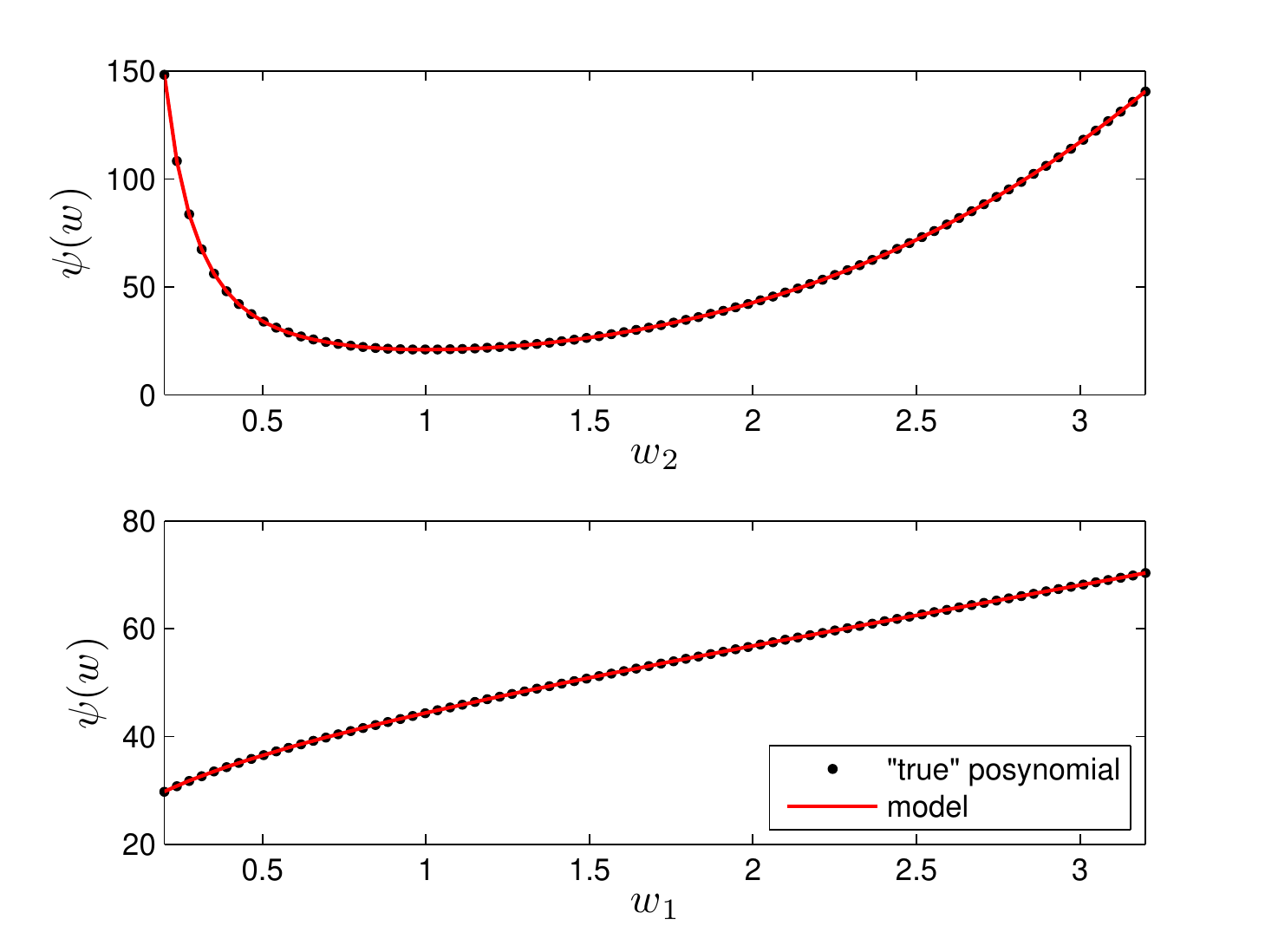}

\caption{Example 1. Comparison between the ``true'' posynomial and the identified
model. Top: section for $w_{1}=2.3$, $w_{3}=0.9$ and $w_{2}\in[0.2,3.2]$.
Bottom: section for $w_{2}=0.7$, $w_{3}=3.1$ and $w_{1}\in[0.2,3.2]$.
\label{sezioni}}

\end{figure}

We set $\lambda_{i}=\gamma\left\Vert \Phi_{i}\right\Vert _{2}^{2}$,
$i=1,\ldots,3294$, and $\sigma=\min_{i}\lambda_{i}/10$. It has been
observed in several numerical experiments that this choice is effective
to penalize monomials with large powers. We considered several values
of $\gamma$, logarithmically spaced in the interval $[10^{-5},10^{-2}]$.
For each value of $\gamma$, the optimization problem (\ref{eq:nnrsqrtLASSO:primal})
has been solved using the approach described in Sections \ref{sec:dual}-\ref{sec:cd}.
Then, the following quantities have been recorded: 
\begin{itemize}
\item the cardinality (i.e., the number of nonzero entries) of the solution
$x$ of the optimization problem (\ref{eq:nnrsqrtLASSO:primal}); 
\item the relative error $RE=\left\Vert \Phi x-y\right\Vert _{2}/\left\Vert y\right\Vert _{2}$. 
\end{itemize}
\noindent Figure~\ref{pareto:figure} shows the Pareto trade-off curve,
reporting the $RE$ values versus the solution cardinality. Based
on this curve, the parameter value $\gamma=10^{-4}$ has been chosen,
since providing the best trade-off between the model complexity (measured
by the cardinality of $x$) and its accuracy (measured by the relative
error $RE$). The model identified with this value of $\gamma$ is
given by
\[
\psi(w)=0.99w_{2}^{1.5}w_{3}^{3}+1.99w_{1}^{2}w_{3}^{-1}+2.94w_{2}^{3.2}+4w_{1}^{0.5}w_{2}^{-2}w_{3}.
\]
It can be noted that the identification algorithm has been able to
recover the ``true'' monomials and to accurately estimate the coefficients
of these monomials. The model is compared with the ``true'' posynomial
in Figure \ref{sezioni}, where some sections of the two functions
are shown.

In order to validate the identified model (supposing that the ``true''
function was not known), a new set of $600$ data has been randomly
generated from (\ref{eq:psi_es2}), where 
the same intervals for $w_{i}$ and the same type of noise and have been considered (although using noise free data for the validation would have allowed
us to assess the quality of the model more accurately, here we used
noise corrupted data to be closer to a real situation). The relative
error obtained by the model on this validation data is $RE=0.012$.

Then, two Monte Carlo simulations have been performed, each consisting
of 100 repetitions of this data generation-identification-validation
procedure. Noise-to-signal standard deviation ratios of 1\% and 3\%
have been considered in the two simulations, respectively. In the
first one, the nnsqrt-LASSO algorithm has been able to find the ``true''
monomials 97\% of times; the average relative error $\bar{RE}=0.011$
on the validation data has been obtained. In the second one, the ``true''
monomials have been recovered 67\% of times; the average relative
error $\bar{RE}=0.021$ on the validation data has been obtained.

We next discuss a few relevant aspects related to the identification
process.

The safe feature elimination discussed in Section \ref{sec:safelim},
reduced the number of columns of $\Phi$ from $3294$ to $2587$ (average
value obtained in the two Monte Carlo simulations), suggesting that
this elimination phase can be useful in practical large-scale problems.

The time taken for applying the safe elimination and solving the optimization
problem (\ref{eq:nnrsqrtLASSO:primal}) with the approach described
in Sections \ref{sec:dual}-\ref{sec:cd} is about $91$ seconds on
a PC with a Core i7 processor and a RAM memory of 8GB (average time
obtained in the two Monte Carlo simulations).

\subsection{Example 2: identification of airfoil drag force}

As a second numerical experiment, we have considered the problem of identifying
a posynomial model for the drag force (per unit length) of a NACA
4412 airfoil. 

This force can be evaluated as a function of the air flow
density $\rho$, the wing chord $\eta$, the incidence angle $\theta$
and the flow velocity $v$, that is 
\[
F_{D}=\psi^{o}(w)
\]
where $w=[\rho\:\eta\:\theta\: v]^{\top}$. No analytical expression is available for this function. The values $\psi^{o}(w)$
can be obtained via simulations based on CFD (computational fluid dynamics),
by integration of the Navier-Stokes equations. Each evaluation is
numerically very costly, thus it is of interest to obtain a simple
model for $F_{D}$, to be used, for instance, in a later stage of
system evaluation or design. 

In this example, we identified a posynomial model for the drag force
of the airfoil, from data obtained from the CFD simulations. The posynomial
form is important since it allows the application of geometric programming
algorithms, which in turn allow for efficient optimization of the
airfoil characteristics, see, e.g., \cite{HoAb12}.

A set $\mathfrak{D}=\left\{ y(k)=\psi^{o}(w(k)),w(k)\right\} _{k=1}^{m}$
of $m=50$ input-output data points has been obtained, for randomly
chosen values of $\rho$, $\eta$, $\theta$ and $v$ in the intervals
shown in Table~\ref{tab_param}. 

\begin{table}[htb]
\centering

\begin{tabular}{|c|c|c|c|}
\hline 
PARAM.  & Minimum  & Maximum  & Dimension \tabularnewline
\hline 
$\rho$  & 0.039  & 1.2250  & $\mathrm{[kg/m^{3}]}$ \tabularnewline
\hline 
$\eta$  & 0.1  & 1  & $\mathrm{[m]}$ \tabularnewline
\hline 
$\theta$  & -5  & 10  & $\mathrm{[deg]}$ \tabularnewline
\hline 
$v$  & 0  & 40  & $\mathrm{[m/s]}$ \tabularnewline
\hline 
\end{tabular}\caption{Parameter intervals considered in the CFD simulations.}

\label{tab_param} 
\end{table}

The exponent sets 
\begin{equation}
Q_{j}=\{-2,-1,0,1,2\},\: j=1,\ldots,4.\label{eq:exp_set}
\end{equation}
have been assumed, following the approach described in Section~\ref{pp_id}.
This choice has been made after a preliminary trial and error process.
For $m=50$ and for the exponent sets (\ref{eq:exp_set}), $\Phi$
results to be a $50\times625$ matrix.

\begin{figure}
\centering
\includegraphics[scale=0.5]{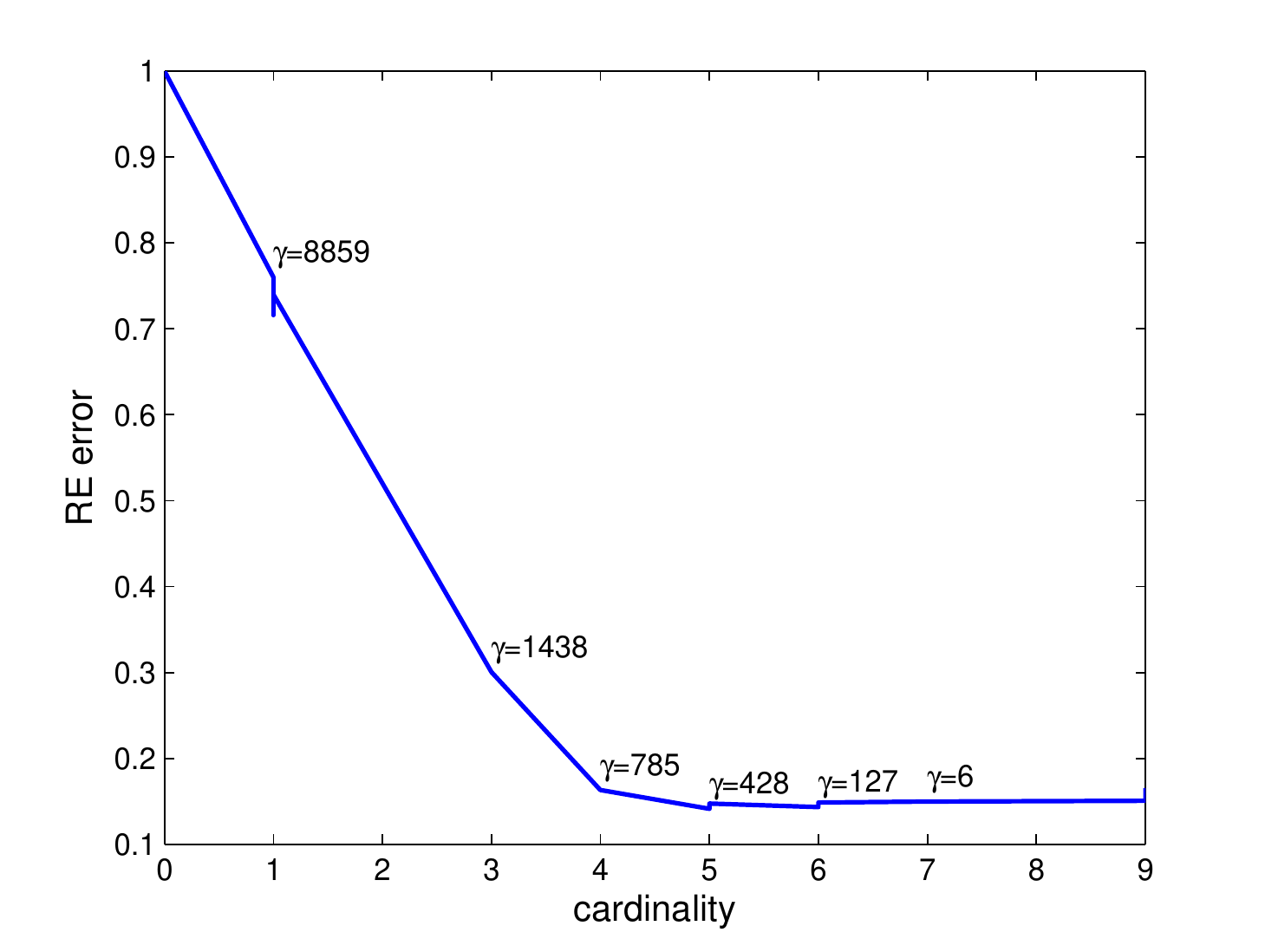}
\caption{Example 2. Pareto trade-off curve.}
\label{pareto:figure2} 
\end{figure}

We set for simplicity $\lam=\gamma\one$, $\sigma=\gamma/10$, and
we considered several values of $\gamma$, logarithmically spaced
in the interval $[1,10^{5}]$. For each value of $\gamma$, the optimization
problem (\ref{eq:nnrsqrtLASSO:primal}) has been solved using the
approach described in Sections~\ref{sec:dual}-\ref{sec:cd}. For
each value of $\gamma$, the following quantities have been recorded:
\begin{itemize}
\item the cardinality (i.e., the number of nonzero entries) of the solution
$x$ of the optimization problem (\ref{eq:nnrsqrtLASSO:primal}); 
\item the relative error $RE=\left\Vert \Phi x-y\right\Vert _{2}/\left\Vert y\right\Vert _{2}$.
\end{itemize}
\noindent Figure~\ref{pareto:figure2} shows the Pareto trade-off curve,
reporting the $RE$ values versus the solution cardinality. Based
on this curve, the parameter value $\gamma=785$ has been chosen,
since providing the best trade-off between the model complexity (measured
by the cardinality of $x$) and its accuracy (measured by the relative
error $RE$).

In order to verify the reliability of an identified model, we carried
out a leave-one-out (LOO) cross validation, on a subset of the available
data. In particular, we used for cross validation data points $w(j)$
that lie within $0.75\%$ from the boundary of the the hyperrectangle
defining the minimum and maximum deviation for each parameter (as
defined in Table \ref{tab_param}). This was done to avoid points
near the boundary of the $w$ domain, which are too close to the non-explored
region.

For each pair $(y(j),w(j))$ in the LOO validation set, a posynomial
model has been identified from the data set $\mathfrak{D}\setminus(\ensuremath{y(j),w(j))}$.
This model has then been tested on the single datum $(y(j),w(j))$,
and the relative error $\nu_{j}=|y(j)-\hat{y}(j)|/\|y_{LOO}\|_{2}$
has been evaluated, where $\hat{y}(j)$ is the output provided by
the model, and $\|y_{LOO}\|_{2}$ is the Euclidean norm of the vector
with entries $y(j)$, for $j$ in the validation set. The accumulated
relative error is given by $AE=\sqrt{\sum_{j}\nu_{j}^{2}}$. In our
experiment, with $\gamma=785$, we obtained $AE=0.25$. This value
appears to be quite low: a model identified using the proposed approach
is able to approximate the unknown function quite accurately, even
if only $50$ points are used to explore its 4-dimensional domain.

The same LOO validation has been performed considering $\gamma=1438$
and $\gamma=127$, obtaining $AE=0.38$ and $AE=0.25$, respectively.
The model identified using $\gamma=785$ has thus the most advantageous
trade-off between complexity and accuracy. This model is given by
\[
\psi(w)=x_{340}\eta v^{2}+x_{440}\rho v^{2}+x_{465}\rho\eta v^{2}+x_{565}\rho^{2}v^{2}
\]
where $x_{340}=1.2746\times10^{-4}$, $x_{440}=3.5469\times10^{-3}$,
$x_{465}=2.8703\times10^{-4}$, and $x_{565}=5.0722\times10^{-4}$
(the units of these coefficient can be inferred from Table \ref{tab_param}).
It is interesting to note that a dependence of the drag force on the
square velocity has been found by the algorithm and this result is
consistent with the well-known drag equation. No significant dependence
on the incidence angle $\theta$ has been observed. A possible interpretation
for this latter result is that the range considered for $\theta$
is not sufficiently large compared to the ranges considered for $\rho$,
$\eta$ and $v$ (see Table \ref{tab_param}) and, consequently, the
force variations due to $\theta$ are negligible with respect to those
produced by the other three parameters.

We next discuss a few relevant aspects related to the identification
process.

The safe feature elimination discussed in Section \ref{sec:safelim},
reduced the number of columns of $\Phi$ from $625$ to $222$ (this
latter is the average value obtained in the LOO validation), suggesting
that this elimination phase can be quite useful in practical large-scale
problems.

The time taken for applying the safe elimination and solving the optimization
problem (\ref{eq:nnrsqrtLASSO:primal}) with the approach described
in Sections \ref{sec:dual}-\ref{sec:cd} is about $0.35$ seconds
on a PC with a Core i7 processor and a RAM memory of 8GB (average
time obtained in the LOO validation).

{\bf Acknowledgments:}
We thank Valentina Dolci (Politecnico di Torino) for providing us
with the fluid dynamic simulation data used in the example.

\section{Conclusions}

An approach for the identification of posynomial models has been presented in this paper, based on the solution of a nonnegative regularized square-root LASSO  problem. In this approach, a large-scale expansion of monomials is considered and the model is identified by seeking coefficients of the expansion that minimize an objective composed by a fitting error term and a sparsity promoting term. A sequential coordinate-descent scheme has been proposed to solve the nnrsqrt-LASSO problem. This scheme guarantees convergence to a minimum of the objective function and is suitable for large-scale implementations. Two numerical examples have finally been shown to demonstrate the effectiveness of the approach. The first one regards identification of a posynomial with negative and non integer exponents; the second one is about identification of a posynomial model for a NACA 4412 airfoil.

\bibliographystyle{agsm}
\bibliography{poly_posy_nomial,references_1,mybiblio}

\end{document}